\documentclass[twocolumn,showpacs,nofootinbib,amsmath,amssymb,floatfix,prd]{revtex4}
\usepackage{mathrsfs}
\usepackage{latexsym}
\usepackage{amsmath}
\usepackage{amsbsy}
\usepackage{amssymb}
\usepackage{epsfig}
\usepackage{graphicx}
\usepackage{dcolumn}
\usepackage{graphicx}
\usepackage{color}

\newcommand{\gt}{\tilde{g}}

\newcommand{\Aether}{\AE ther }

\newcommand{\varphio}{\mathring{\varphi}}
\newcommand{\gto}{\mathring{\gt}}

\newcommand{\al}{\alpha}
\newcommand{\be}{\beta}
\newcommand{\da}{\partial}
\newcommand{\om}{\omega}
\newcommand{\de}{\delta}

\begin{document}

\title{Stability and Quasinormal Modes of Black holes in Tensor-Vector-Scalar theory: Scalar Field Perturbations}

\author{Paul D. Lasky}
    \email{lasky@tat.physik.uni-tuebingen.de}
    \affiliation{Theoretical Astrophysics, Eberhard Karls University of T\"ubingen, T\"ubingen 72076, Germany}
\author{Daniela D. Doneva}
    \email{ddoneva@phys.uni-sofia.bg}
    \affiliation{Department of Astronomy, Faculty of Physics, St. Kliment Ohridski University of Sofia, 1164 Sofia, Bulgaria}
    \affiliation{Theoretical Astrophysics, Eberhard Karls University of T\"ubingen, T\"ubingen 72076, Germany}

        \begin{abstract}
    The imminent detection of gravitational waves will trigger precision tests of gravity through observations of quasinormal ringing of black holes.  While General Relativity predicts just two polarizations of gravitational waves, the so-called plus and cross polarizations, numerous alternative theories of gravity predict up to six different polarizations which will potentially be observed in current and future generations of gravitational wave detectors.  Bekenstein's Tensor-Vector-Scalar (TeVeS) theory and its generalization fall into one such class of theory that predict the full gamut of six polarizations of gravitational waves.  In this paper we begin the study of quasinormal modes (QNMs) in TeVeS by studying perturbations of the scalar field in a spherically symmetric background.  We show that, at least in the case where superluminal propagation of perturbations is not present, black holes are generically stable to this kind of perturbation.  We also make a unique prediction that, as the limit of the various coupling parameters of the theory tend to zero, the QNM spectrum tends to $1/\sqrt{2}$ times the QNM spectrum induced by scalar perturbations of a Schwarzschild black hole in General Relativity due to the intrinsic presence of the background vector field.  We further show that the QNM spectrum does not vary significantly from this value for small values of the theory's coupling parameters, however can vary by as much as a few percent for larger, but still physically relevant parameters.
        \end{abstract}

        \received{\today}\published{}



        \maketitle

\section{Introduction}

Perturbations of black holes have been intensely studied during the past decades with relation to black hole stability, astrophysical implications of gravitational wave detection and, more recently, gauge/gravity dualities (see the reviews \cite{kokkotas99,nollert99,berti09,konoplya10}).  From an astrophysical perspective, the impending observation of gravitational waves from black holes will herald new and unprecedented tests of gravity through two main avenues.  The first is through the possible detection of gravitational wave polarizations that are {\it not} predicted by General Relativity (GR) such as scalar ``breathing modes'' and extra tensor and vector degrees of freedom\footnote{See \cite{eardley73a,eardley73b} for possible gravitational modes in metric theories of gravity, and \cite{will93,will06} as well as \cite{bianchi96,bianchi98,bassan10} and references therein for discussions of possible detections with inteferometer and spherical gravitational wave detectors resepctively.}.  The second method is a direct probe of the no-hair theorem of GR through observations of the quasinormal ringing of black holes known as quasinormal modes (QNMs).  General Relativity predicts that the simultaneous measurement of the frequency and damping time of a single QNM is sufficient to determine both the mass and angular momentum of a black hole \cite{echeverria89,finn92,flanagan98,dreyer04,berti06}, implying the detection of more than one QNM from a single black hole can give rigorous constraints on the no-hair theorem -- one of the major science goals for ground based and space based gravitational wave detectors \cite[e.g.][]{berti06,goggin06,berti07,yunes09}.  An alternative approach to the measurement of the frequencies of black hole QNMs based on a connection between the emission of gravitational waves and strong gravitational lensing has recently been elucidated in \citet{stefanov10}.

A recently popular alternative theory of gravity that predicts both extra polarizations of gravitational wave propagation as well as a violation of the no-hair theorem is Bekenstein's Tensor-Vector-Scalar (TeVeS) theory \cite{bekenstein04} and its generalization \cite{contaldi08,skordis08,sagi09}.  In this paper we take a first step towards exploring the perturbations of black holes within TeVeS and its generalization by studying perturbations of the background scalar field.  This enables us to look at both the stability of black holes in the theories, as well as looking at how the QNM frequencies and damping times are affected by the various parameters of the theories.

TeVeS is a covariant generalization of the Modified Newtonian Dynamics (MoND) paradigm \cite{milgrom83}, which attempts to explain the discrepancy between observed and predicted mass distributions on galactic scales.  As a tool for interpreting galactic scale observations, MoND has proven extremely successful (see \cite{sanders02} for a review).  However, as a non-covariant theory it is ill-equipped to explain the full gamut of astrophysical and cosmological observations.  To this end, TeVeS has recently been employed in an attempt to explain, amongst other things, gravitational lensing \cite{chiu06,chen06,shan08,chiu08,feix10}, the cosmic microwave background power spectrum \cite{dodelson06,angus09}, large-scale cluster surveys \cite{skordis06} and even type Ia supernova \cite{diazrivera06,zhao07}, while retaining the successful predictions of MoND in the weak acceleration limit (see \cite{skordis09} for a recent review of TeVeS).

In the Newtonian regime, TeVeS has been shown to reproduce the parametrized post-Newtonian coefficients to a level consistent with solar system experiments \cite{bekenstein04}.  In the strong field regime however, comparatively little has been studied.  \citet{giannios05} first solved the field equations for static, spherically symmetric, vacuum spacetimes within TeVeS.  He found two branches of solutions dependent on the degrees of freedom allowed in the vector field.  Giannios' solution was, however, plagued by the fact that the scalar field was necessarily negative for some radii, implying superluminal propagation of scalar waves \cite{bekenstein04}.  \citet{sagi08} remedied this situation by showing that another branch of solutions to the field equations exist that has the same physical metric, but where the scalar field is positive throughout the spacetime.  \citet{sagi08} also found the charged Reissner-Nordstr\"om solution within TeVeS and provided a detailed study of black hole thermodynamics.  \citet{lasky08} then found the Tolman-Oppenheimer-Volkoff solution in order to study neutron star structure, work which has since been extended to include slow rotation \cite{sotani10} and fluid \& spacetime perturbations \cite{sotani09,sotani09a}.

Meanwhile, trouble was brewing with the original formulation of the TeVeS field equations.  \citet{seifert07} first showed that the Schwarzschild-TeVeS solution is unstable to linear perturbations for experimentally and phenomenologically valid values of the various coupling parameters.  \citet{contaldi08} then showed that the vector field is prone to the formation of caustics in a variety of simple dynamical situations, analogously to Einstein-\Aether theories where the vector field is described by a Maxwellian action.  Finally, \citet{sagi09} showed that the vector field is constrained by the cosmological value of the scalar field in such a way that it prevents the scalar field from evolving.

To overcome the aforementioned issues, \citet{contaldi08}, \citet{skordis08} and \citet{sagi09} independently provided a generalization of TeVeS by taking a more general action for the vector field which is motivated by Einstein-\Aether theory.  Complicating an already complicated theory is a complicated process.  It has implied that little rigorous work has been achieved looking at the structure of the field equations.  \citet{skordis08} studied the cosmological equations of the theory and showed they are identical to the equations governing the original TeVeS cosmology up to a rescaling of Hubble's constant.  \citet{sagi09} looked at the parametrized post-Newtonian (PPN) parameters of the theory and found them not to be in conflict with solar system experiments for suitable values of the coupling parameters.

It should be mentioned that the generalized TeVeS theory is also not without its problems.  In two papers, \citet{mavromatos09} and \citet{ferreras09} have shown that observations of gravitational lensing and galactic rotation curves provide an inconsistent parameter space for the theory.  Although their results were based on the original TeVeS theory, their formulation of the problem carries over to the generalized theory based on their use of a vector field with only a temporal component (the proof of this is the same as that given in \citet{lasky09} in the strong-field regime).  However, it is exactly this point that diminishes the robustness of their claims.  One main driving force for introducing the vector field into the theory is to provide sufficient gravitational lensing without requiring dark matter \cite{bekenstein04}.  It is therefore not surprising that suppressing the degrees of freedom of this vector field by setting all spatial components to zero also suppresses the degree to which gravitational lensing observations can be made to be consistent with other observations.  With the extra complexity of the vector field in the generalized theory, combined with the relatively primitive state of parameter space estimation, one would expect that gravitational lensing observations could be induced into conforming with galactic rotation curve  observations with a more robust form of the vector field.

The study of gravitational wave propagation in Generalized TeVeS was first specifically broached by \citet{sagi10}, who studied linear perturbations of the field equations to look at the speed and form of gravitational wave propagation.  Sagi showed that there exist six different modes of gravitational waves (as opposed to two in GR) which propagate at four different speeds, all different to the speed of light.  As expected, the various propagation speeds of the gravitational wave modes were found to be dependent on the various coupling parameters in the theory.  It is interesting to note that this is seemingly in contrast with the result of \citet{kahya07} and \citet{desai08} who predict gravitational waves will propagate at the same speed independently of the coupling parameters.  The obvious discrepancy between these two results implies further investigation is necessary.

Spherically symmetric vacuum, charged and perfect fluid solutions of the generalized TeVeS field equations were studied in \citet{lasky09} where the vector field was assumed to contain only a temporal component.  Under these symmetry assumptions it was shown that these solutions are the same as for the original TeVeS theory by \citet{giannios05}, \citet{sagi08} and \citet{lasky08} respectively, where only a rescaling of the vector field coupling parameters is required.  That is, given the aforementioned rescaling, the form of the background scalar, vector and tensor fields, and therefore also the form of the physical metric, are exactly the same in the two theories.  In \citet{lasky09} it was further shown that this result is {\it not} generalizable -- solutions with time dependence or more complicated geometry in the vector or tensor fields will be different between the two theories.  As we are dealing in the present paper with perturbations of the spherically symmetric vacuum solution, we are dealing with the solution of both the original and generalized versions of TeVeS.  Moreover, as the difference between these two theories is in the vector field, the perturbations we perform of the scalar field also hold for both theories.  That is, all work presented in this paper is applicable to both the original version and generalized version of TeVeS.  As such, unless explicitly mentioned, we herein refer to TeVeS to mean the all-encompassing generalized version of the theory.

TeVeS admits the Schwarzschild solution as a possible geometry \cite{giannios05,lasky09}, and most likely the Kerr solution as is the case with a majority of alternative theories of gravity \cite{psaltis08a} (although we note there exists various theories of gravity that do not admit the Kerr solution, for example Chern-Simons gravity \cite{yunes09a}).  Therefore, electromagnetic observations of the spacetime surrounding black holes may not allow for the distinction between GR, TeVeS and alternative theories.  For example, observing the dynamics of stars orbiting close to Sgr A$^{\star}$ will soon be yielding fruitful information about the spacetime structure of that black hole \cite{will08,merritt09}.  This problem however, is well approximated by the stars acting like zero-mass test particles, implying one is only probing geodesics of the background spacetime.  Therefore, measuring the spacetime to be a Kerr black hole using this method does not rule out alternative theories of gravity that admit Kerr as a possible geometry.  The perturbations of such spacetimes (relevant for gravitational waves) however, depends on the specific field equations of the theory, and hence will differ between theories.  This implies QNMs of various Kerr (or Schwarzschild) geometries behave differently and can be used to provide rigorous tests of the theory of gravity.

The article is set out as follows; in section \ref{GenEquations} we provide a brief primer on the relevant field equations for the generalized TeVeS theory, reviewing the structure of the background spacetime in section \ref{sssvac}.  In section \ref{pert} we derive the wave equation governing scalar field perturbations and also discuss analytic results in the limit of small couplings.  In section \ref{TimeEvol}, we analyse the stability of the black hole solutions and then in section \ref{Freq} we compute the QNM spectrum.  Throughout the article Greek indices range from $0\ldots3$ and antisymmemetrization and symmetrization are respectively denoted by square and round brackets: $A_{\left[\mu\nu\right]}:=A_{\mu\nu}-A_{\nu\mu}$, $A_{\left(\mu\nu\right)}:=A_{\mu\nu}+A_{\nu\mu}$.

\section{Generalized TeVeS equations}\label{GenEquations}
TeVeS is built upon three dynamical fields; the Einstein metric, $g_{\mu\nu}$, a time-like normalized vector field, $A^{\mu}$, and a dynamical scalar field, $\varphi$.  A physical metric, $\gt_{\mu\nu}$, in which clocks and rulers are measured, is related to the other three fields according to
\begin{align}
    \tilde{g}_{\mu\nu}={\rm e}^{-2\varphi}\left(g_{\mu\nu}+A_{\mu}A_{\nu}\right)-{\rm e}^{2\varphi}A_{\mu}A_{\nu}.\label{physicalmetric}
\end{align}
This implies that variations in the tensor, vector or scalar fields can be measured through their coupling to the physical metric.

The modified Einstein field equations are
\begin{align}
    G_{\mu\nu}=8\pi G\left[\tilde{T}_{\mu\nu}+\left(1-{\rm e}^{-4\varphi}\right)A^{\al}\tilde{T}_{\al(\mu}A_{\nu)}+\tau_{\mu\nu}\right]+\Theta_{\mu\nu},\label{Einstein}
\end{align}
where $G_{\mu\nu}$ is the Einstein tensor associated with the Einstein frame, $\tilde{T}_{\mu\nu}$ is the physical stress-energy tensor and $\tau_{\mu\nu}$ and $\Theta_{\mu\nu}$ are the effective stress-energy terms associated with the scalar and vector fields respectively.  In particular
\begin{align}
    \tau_{\mu\nu}:=&\frac{\mu}{kG}\Big[\nabla_{\mu}\varphi\nabla_{\nu}\varphi-\frac{1}{2}g^{\al\be}\nabla_{\al}\varphi\nabla_{\be}\varphi g_{\mu\nu}\notag\\
    &-A^{\al}\nabla_{\al}\varphi\left(A_{(\mu}\nabla_{\nu)}\varphi-\frac{1}{2}A^{\be}\nabla_{\be}\varphi g_{\mu\nu}\right)\Big]\notag\\
    &-\frac{\mathcal{F\left(\mu\right)}}{2k^{2}\ell^{2}G}g_{\mu\nu},
\end{align}
where $k$ is the scalar field coupling constant, $\mu$, a function of the theory's coupling parameters and the scalar field, is associated with the MoND acceleration scale, $\mathcal{F}\left(\mu\right)$ is related to the interpolation function in MoND and hence is not {\it a priori} predicted by the theory and $\ell$ is a fixed length scale associated with the free function.  For the remainder of this article we only consider the strong-field regime of the theory, whereby $\mu=1$ is an excellent approximation (for more details see
\cite{bekenstein04,giannios05,sagi08,contaldi08}).  In this case, \citet{contaldi08} showed that the function $\mathcal{F}$ logarithmically diverges, but is exactly cancelled in the field equations, implying this function has zero contribution in this limit.  Moreover, it has been shown \cite{sagi08} that this limit of the strong-field approximation, whereby we allow $\mu=1$, is correct out to at least a million times the gravitational radius off a black hole.  Not only is this well into the asymptotically flat region of the black hole, but for the purpose of QNM calculations this is more than sufficient for the external boundary condition.  That is, it is sufficient in an astrophysical sense, that when we are treating QNMs and their boundary conditions at ``infinity'', we are still dealing with the strong-field limit of the theory.

One key difference between the original version of TeVeS developed by \citet{bekenstein04} and the current version with Maxwellian action \cite{contaldi08,skordis08} is the introduction of three extra vector coupling constants which describe the relative strengths of the individual terms in the vector field action.  The four vector field coupling constants are denoted by $K$, $K_{+}$, $K_{2}$ and $K_{4}$, where the original theory is regained by setting the last of these three parameters to zero.  The vector field contribution to the effective stress-energy in the modified Einstein equations, $\Theta_{\mu\nu}$, can now be expressed as a sum of terms associated with each of the vector field coupling constants, plus a term associated with the Lagrange multiplier, $\lambda$ (which ensures normalization of the vector field);
\begin{align}
    \Theta_{\mu\nu}:=\Theta^{K}_{\mu\nu}+\Theta^{K_{+}}_{\mu\nu}+\Theta^{K_{2}}_{\mu\nu}+\Theta^{K_{4}}_{\mu\nu}+\Theta^{\lambda}_{\mu\nu},
\end{align}
where
\begin{align}
    \Theta^{K}_{\mu\nu}:=&K\left(F_{\al\mu}{F^{\al}}_{\nu}-\frac{1}{4}g_{\mu\nu}F_{\al\be}F^{\al\be}\right),
\end{align}
\begin{align}
    \Theta^{K_{+}}_{\mu\nu}:=&K_{+}\Big[S_{\mu\al}{S_{\nu}}^{\al}-\frac{1}{4}g_{\mu\nu}S_{\al\be}S^{\al\be}\notag\\
        &+\nabla_{\al}\left(A^{\al}S_{\mu\nu}-{S^{\al}}_{(\mu}A_{\nu)}\right)\Big],
\end{align}
\begin{align}
    \Theta^{K_{2}}_{\mu\nu}:=&K_{2}\Big[g_{\mu\nu}\nabla_{\al}\left(A^{\al}\nabla_{\be}A^{\be}\right)-A_{(\mu}\nabla_{\nu)}\left(\nabla_{\al}A^{\al}\right)\notag\\
        &-\frac{1}{2}g_{\mu\nu}\nabla_{\al}A^{\al}\nabla_{\be}A^{\be}\Big],
\end{align}
\begin{align}
    \Theta^{K_{4}}_{\mu\nu}:=&K_{4}\Big[\dot{A}_{\mu}\dot{A}_{\nu}+\dot{A}_{\al}A_{(\mu}\nabla_{\nu)}A^{\al}-\nabla_{\al}\left(\dot{A}^{\al}A_{\mu}A_{\nu}\right)\notag\\
        &-\frac{1}{2}g_{\mu\nu}\dot{A}_{\al}\dot{A}^{\al}\Big],
\end{align}
\begin{align}
    \Theta^{\lambda}_{\mu\nu}:=&-\lambda A_{\mu}A_{\nu}.\label{ThetaLambda}
\end{align}
Here, $F_{\mu\nu}:=\nabla_{[\nu}A_{\mu]}$, $S_{\mu\nu}:=\nabla_{(\nu}A_{\mu)}$ and $\dot{A}_{\mu}:=A^{\al}\nabla_{\al}A_{\mu}$.

Variation of the total action with respect to the vector field gives the vector field equation
\begin{align}
    K&\nabla_{\al}F^{\mu\al}+K_{+}\nabla_{\al}S^{\al\mu}+K_{2}\nabla^{\mu}\left(\nabla_{\al}A^{\al}\right)+\lambda A^{\mu}\notag\\
        &+K_{4}\left[\nabla_{\al}\left(\dot{A}^{\mu}A^{\al}\right)-\dot{A}^{\al}\nabla^{\mu}A_{\al}\right]+\frac{8\pi\mu}{k} A^{\al}\nabla_{\al}\varphi g^{\mu\be}\nabla_{\be}\varphi\notag\\
        =&8\pi G\left(1-{\rm e}^{-4\varphi}\right)g^{\mu\al}\tilde{T}_{\al\be}A^{\be}.\label{vector}
\end{align}
Contraction of this equation with the vector field isolates the Lagrange multiplier.  This subsequent equation can then be used in the modified Einstein equation, in particular in the term expressed in equation  (\ref{ThetaLambda}), such that the system is fully determined.

The final field equation is that of the scalar field
\begin{align}
    \nabla_{\be}&\left[\mu\left(g^{\al\be}-A^{\al}A^{\be}\right)\nabla_{\al}\varphi\right]\notag\\
        &=kG\left[g^{\al\be}+\left(1+{\rm e}^{-4\varphi}\right)A^{\al}A^{\be}\right]\tilde{T}_{\al\be}.\label{scalar}
\end{align}

It is of interest for the current work how the TeVeS field equations limit to the GR equations.  This is achieved by continuously limiting $k$ and all of the $K$'s to zero, as well as taking $\ell\rightarrow\infty$.  This was shown in detail for the original TeVeS theory by \citet{bekenstein04}, his sections III C and D.

\section{Spherically Symmetric, Static, Vacuum Solutions}\label{sssvac}
In order to discuss perturbations of black holes, it is worth spending some time discussing the current status of spherically symmetric, static, vacuum solutions in TeVeS.  \citet{giannios05} first solved the original TeVeS equations of Bekenstein for the case of a spherically symmetric, static, vacuum spacetime.  He showed that there exist two branches of solutions based on the form of the vector field; one branch where the vector field is temporally aligned
and another when the vector field has both a temporal and radial component.
\citet{lasky09} then found the equivalent solution in the case of a purely temporal vector field for the more general TeVeS theory.  In that work it was shown that the solution in the generalized TeVeS theory is the same as that in the original TeVeS theory with a simple substitution of the parameters of the theory

Following the notation of \citet{giannios05}, we write the physical line element in isotropic coordinates in terms of two constants, $r_{c}$ and $a$;
\begin{align}
    d\tilde{s}^{2}=&-\left(\frac{1-r_{c}/r}{1+r_{c}/r}\right)^{a}dt^{2}\notag\\
        &+\left(1+\frac{r_{c}}{r}\right)^{2+a}\left(1-\frac{r_{c}}{r}\right)^{2-a}\left(dr^{2}+r^{2}d\Omega^{2}\right),\label{metric}
\end{align}
where $d\Omega^{2}:=d\theta^{2}+\sin^{2}\theta d\phi^{2}$.  The two constants in the physical metric are related to the various coupling parameters of the theory, including the ``scalar mass'', $m_{\varphi}$, as well as the characteristic gravitational radius, $r_{g}$, according to
\begin{align}
    r_{c}&=\frac{r_{g}}{4}\sqrt{1+\frac{k}{\pi}\left(\frac{Gm_{\varphi}}{r_{g}}\right)^{2}-\frac{\mathcal{K}}{2}},\label{rc}\\
    a&=\frac{r_{g}}{2r_{c}}+\frac{kGm_{\varphi}}{4\pi r_{c}},\label{a}
\end{align}
where $\mathcal{K}:=K+K_{+}-K_{4}$. Note that all quantities in equation (\ref{a}) are greater than or equal to zero, implying $a\ge0$.  Moreover, there are only two situations where $a$ can be identically zero.  Firstly, if both the gravitational radius and the scalar field coupling constant, $r_{g}$ and $k$, both vanish.  Not only does this reduce the theory to the Einstein-\Aether theory, but it also implies that the metric becomes Minkowski.  The second possibility is that $r_{c}$ diverges, which is unphysical.  It is therefore only possible that $a$ is strictly greater than zero.

Equation (\ref{metric}) describes more than just a black hole spacetime.   \citet{giannios05} showed that $r_{c}$ is a black hole event horizon if and only if $a=2$.  This is based on the requirement that the surface at $r=r_{c}$ must have a finite surface area, and also the singularity must be removable.  Evaluating the physical Ricci scalar shows that it diverges for values of $a<2$ and also $2<a<4$.  Meanwhile, as the surface area is proportional to the $\gt_{rr}(r_{c})$ component of the metric, one finds that this diverges for $a>2$.  One therefore has a black hole solution only when $a=2$.  When $0<a<2$ and $2<a<4$ one has a naked singularity at $r=r_{c}$, and $a\ge4$, $r_{c}$ represents a removable singularity with a divergent surface area.

As mentioned, the only solution given by the metric (\ref{metric}) that represents a black hole is that with $a=2$.  One immediately notes that this is exactly equivalent to the Schwarzschild metric in isotropic coordinates.

\subsection{Equivalence with Brans-Dicke Theory}
It is further interesting to note that the metric of equation (\ref{metric}) is exactly equivalent to the Brans `type I' metric \cite{brans62} in Brans-Dicke scalar tensor theory \cite{brans61} (with the association $a=2Q$ and $\chi=0$ from the notation of \citet{scheel95b}) \cite{lasky10}.  Note that in the Brans I solution, setting $\chi=0$ necessarily implies that the parameter $Q$ is unity, and hence the Schwarzschild solution is recovered.  This is because the Brans I solution has an extra algebraic equation which links $Q$ and $\chi$.

A popular representation of the Brans type I metric can be found by taking the coordinate transformation
 given by
 \begin{align}
    R(r)=r\left(1+r_{c}/r\right)^{2},\label{SchwCoord}
\end{align}
from the metric (\ref{metric}) and defining $M:=2r_{c}$, one gets
\begin{align}
    d\tilde{s}^{2}=&-\left(1-\frac{2M}{R}\right)^{a/2}dt^{2}+\left(1-\frac{2M}{R}\right)^{-a/2}dR^{2}\notag\\
    &+R^{2}\left(1-\frac{2M}{R}\right)^{1-a/2}d\Omega^{2}.\label{BDmetric}
\end{align}
In particular, this is the line element expressed as equation (7) of \citet{campanelli93} with the association $m=-n=a/2-1$.  For a majority of the article we will deal with the case of $a=2$, for which the line element (\ref{BDmetric}) becomes exactly the Schwarzschild line element.

The association with the Brans-Dicke theory is interesting given the vast complexity of the field equations in TeVeS as compared with Brans-Dicke theory.  However, the association is due to the simplifying assumption that the vector field is purely timelike.  It was shown for the original version of TeVeS that black hole solutions with a non-zero radial component for the vector field produce significantly more complicated geometries than the case where the radial component vanishes \cite{giannios05}.  This complexity will only increase when one analyses the black hole solutions with non-zero radial vector fields in the fully general theory.

It is clear that, while the background spacetime in TeVeS and Brans-Dicke theories are equivalent, perturbations of these spacetimes will differ as the scalar field equation of TeVeS is {\it not} a Klein-Gordon equation but includes contributions from the background vector field.  In fact, whilst scalar field perturbations of Schwarzschild-Brans-Dicke black holes [equation (\ref{BDmetric}) with $a=2$] have the same QNM spectrum as that of scalar perturbations of Schwarzshild black holes in GR\footnote{This was not explicitly shown in \citet{kwon86} who analyzed the stability of such perturbations, however the wave equation they derive is equivalent to the Regge-Wheeler wave equation for scalar perturbations of Schwarzschild black holes in GR.} we shall show in section \ref{Freq} that perturbations in TeVeS have a different spectrum.


\subsection{Vector and Scalar Field Parameters}
The above metric (\ref{metric}) is that found by \citet{giannios05}.  He displayed trepidation towards this solution because the scalar field becomes negative close to the horizon, which therefore allows for superluminal propagation of scalar waves \cite{bekenstein04}.  \citet{sagi08} overcame this issue by showing that \citet{giannios05} overlooked a branch of solutions where the scalar field was everywhere positive, although they did this only for the case where $a=2$.  Using equations (\ref{rc}) and (\ref{a}) we can re-derive the extra solution of \citet{sagi08} with the generalization that $a$ remains arbitrary.  In this case, one can show that the scalar field throughout the spacetime is given by
\begin{align}
    \varphi(r)&=\varphi_{c}+\delta_{\pm}\ln\left(\frac{1-r_{c}/r}{1+r_{c}/r}\right)\notag\\
        &=\varphi_{c}+\delta_{\pm}\ln\left(\sqrt{1-\frac{2M}{R}}\right),\label{varphi}
\end{align}
where $\varphi_{c}$ is the cosmological value of the scalar field, which can be determined through cosmological observations, and $\delta_{\pm}$ can be found to be
\begin{align}
    \delta_{\pm}:=\frac{a\left(2-\mathcal{K}\right)k/2\pm\sqrt{2k\left[\left(k-a^{2}\pi\right)\left(2-\mathcal{K}\right)+8\pi\right]}}{\left(2-\mathcal{K}\right)k+8\pi}.\label{delta}
\end{align}
Note here that when $a=2$, equation (\ref{delta}) reduces to exactly equation (67) of Ref. \cite{sagi08} (where $\mathcal{K}\equiv K$ as they were working in the original TeVeS which has $K_{+}=K_{2}=K_{4}=0$).

Given that $0<\mathcal{K}<2$ and $k>0$, one can trivially show that $\delta_{+}>0$.  Moreover, after some extra work a conditions is derived for $\delta_{-}$ that is
\begin{align}
a^{2}<\frac{8}{2-\mathcal{K}}\qquad\Longrightarrow\qquad\delta_{-}<0\label{ineqaK}
\end{align}
which further implies that $\delta_{-}<0$ for all $a\le2$.  A majority of the results of this paper are for the cases where $a=2$, which is the only black hole solution of the field equations.  In this case it is clear that $\delta_{-}$ is always negative, a fact that will be important for the perturbation analysis in section \ref{BHperts}.

It is interesting to note that there exists a special case of the above functions whereby $\delta_{-}$ vanishes identically, implying that $\varphi(r)=\varphi_{c}$ throughout the spacetime.  It is apparent that this is only the case where $a^{2}=8/(2-\mathcal{K})$ (or when $\mathcal{K}=0$ which is not of interest for the present work).  Whilst this case exhibits a constant scalar field throughout the spacetime, there still exists an irremovable naked singularity at $r=r_{c}$.  Although the stability of such objects is of great interest, we shall see below that the wave equation governing the perturbations for this case is no simpler than the generic case when there is a naked singularity at the horizon.  As such, we leave this for further exploration in a latter paper.

The generic behaviour of the scalar field throughout the spacetime is of great importance to the present discussion due to the presence of superluminal propagation of scalar waves.  Indeed, \citet{bekenstein04} showed that, in the eikonal approximation, scalar perturbations in the Newtonian limit of the theory travel with velocity $v=\exp\left(-2\varphi\right)/\sqrt{2}$.  This implies that scalar perturbations travel superluminally if and only if $\varphi<0$.  Equation (\ref{varphi}) shows that the scalar field necessarily diverges as $r\rightarrow r_{c}$.  When $\delta_{\pm}<0$, the scalar field diverges to positive infinity, and indeed the scalar field is everywhere positive (given $\varphi_{c}>0$).  However, when $\delta_{\pm}>0$, the scalar field diverges to negative infinity as one approaches the critical radius, $r_{c}$, implying for some values of the radial coordinate the scalar field is negative.

Finally, for the remaining sections we will require the form of the vector field in the prescribed background spacetime, which can be shown to be
\begin{align}
    A^{t}&={\rm e}^{\varphi}\left(\frac{1-r_{c}/r}{1+r_{c}/r}\right)^{-a/2}\notag\\
        &={\rm e}^{\varphi}\left(1-\frac{2M}{R}\right)^{-a/4}.
\end{align}
We note that the various coupling parameters of the theory are mainly expressed here in the ${\rm e^{\varphi}}$ term [see equation (\ref{varphi})].

\section{Perturbation Equations}\label{pert}
We express the perturbed scalar field as
\begin{align}
    \varphi=\varphio+\de\varphi,\qquad{\rm where}\qquad\frac{\de\varphi}{\varphio}<<1,
\end{align}
where $\varphio$ and $\delta\varphi$ are respectively the background and perturbed scalar fields.  Since the scalar field perturbations decouple from perturbations of the vector and tensor fields \cite{sagi10}, we are free to let perturbations of the vector and tensor fields vanish, i.e. $\de A^{\mu}=\de g^{\mu\nu}=0$, throughout the article, and we leave the analysis of these perturbations to a future article.

The perturbation of equation (\ref{physicalmetric}) implies
\begin{align}
    \de\gt_{\mu\nu}=-2\de\varphi\left[{\rm e}^{-2\varphio}\left(g_{\mu\nu}+A_{\mu}A_{\nu}\right)+{\rm e}^{2\varphio}A_{\mu}A_{\nu}\right].
\end{align}
Given this equation, together with the fact that the physical metric of the background spacetime is diagonal, this can be expressed component-wise as
\begin{align}
    \de\gt_{tt}=2\de\varphi\gto_{tt}\qquad{\rm and}\qquad \de\gt_{ii}=-2\de\varphi\gto_{ii}.\label{pertsc}
\end{align}

From the above form of the perturbed physical metric it is obvious that a perturbation of the scalar field only affects the diagonal components of the line element.  These are  the so-called \textit{breathing modes} of gravitational waves which are not present in GR.  The detection of these modes is possible using current interferometric techniques, although requires multiple suitably oriented detectors to be operating simultaneously (see Ref. \cite{will06} and references therein), spherical resonant-mass detectors (for example \cite{bianchi96,bianchi98,maggiore00,coccia00,bassan10}), or through the difference in expected energies between the standard plus and cross polarizations in standard interferometric detection.  The detection of such modes would instantly falsify GR, however not detecting these modes will simply act to constrain both sources and parameter spaces in various theories.

The scalar field equation (\ref{scalar}) is simple to perturb, and turns out to be
\begin{align}
    \nabla_{\be}\left[\left(g^{\al\be}-A^{\al}A^{\be}\right)\da_{\al}\de\varphi\right]=0.\label{scalarwave}
\end{align}

For the remainder of the article we work only with the black hole case in which the metric parameter $a=2$.  In the Appendix we show the more general form of the wave equation with arbitrary $a$, however we leave the study of the stability of such spacetimes to future work.

\subsection{The wave equation}\label{wavesec}
Performing the standard seperation of variables in Schwarzschild coordinates
\begin{align}
    \delta\varphi\left(t,R,\theta,\phi\right)=\frac{\Phi(t,R)}{R}\sum_{\ell=0}^{\infty}\sum_{m=-\ell}^{\ell} Y_{\ell m},
\end{align}
where the $Y_{\ell m}$ are the standard spherical harmonics, leads to the wave equation
\begin{align}
    -2{\rm e}^{4\varphio}\frac{\da^{2}\Phi}{\da t^{2}}+\frac{\da^{2}\Phi}{\da R_{\star}^{2}}-\left(1-\frac{2M}{R}\right)\left[\frac{\ell\left(\ell+1\right)}{R^{2}}+\frac{2M}{R^{3}}\right]\Phi=0,\label{timeevolSchw}
\end{align}
where the standard tortoise coordinate, $R_{\star}(R)$, has been defined according to
\begin{align}
    \frac{dR}{dR_{\star}}=1-\frac{2M}{R}.
\end{align}
An important aspect of this wave equation (\ref{timeevolSchw}) is that it is almost identical to the wave equation governing scalar perturbations of a Schwarzschild black hole in GR, the difference being the factor of $2\exp\left(4\varphio\right)$ preceding the second-order time derivative.  In the weak field limit (i.e. taking large $R$) equation (\ref{timeevolSchw}) reduces to a wave equation where the velocity of the perturbations is $\exp\left(-2\varphio\right)/\sqrt{2}$, in agreement with the velocity of scalar perturbations derived by \citet{bekenstein04} and \citet{sagi10}.

Consider now the limit of equation (\ref{timeevolSchw}) as the background scalar field goes to zero.  This limit is not well defined in the sense that the background scalar field necessarily diverges as $R\rightarrow2M$ for $\delta_{\pm}\neq0$.  However, if we take the limit as $\mathcal{K}$, $k$ and $\varphi_{c}$ tend towards zero, we find the background scalar field tends to zero for all values of $R\neq 2M$.  In this limit we see that the coefficient of the $\da^{2}/\da t^{2}$ term tends to two, which remains distinct from the general relativistic limit.  Consider now a rescaling of the temporal coordinate $t=\sqrt{2}t'$, which acts to reduce the wave equation to exactly the Regge-Wheeler equation for scalar perturbations of Schwarzschild black holes in GR.  If we now follow standard methods and assume a harmonic time dependence for the perturbations
\begin{align}
    \Phi(t,R)\sim{\rm e}^{i\om t} = {\rm e}^{i\om_{s}t'},
\end{align}
where $\om$ and $\om_{s}$ are the QNMs for our TeVeS black hole and the GR Schwarzschild black hole respectively, then we find the following simple relation
\begin{align}
    \lim_{\mathcal{K},k,\varphi_{c}\rightarrow0}\om=\frac{\om_{s}}{\sqrt{2}}\label{root2}
\end{align}
That is, as the theory's various coupling parameters tend to zero, the QNM frequencies and damping times are related to the Schwarzschild QNM frequencies and damping times through a factor of $\sqrt{2}$.  We note again that the GR Schwarzschild black hole spectrum is equivalent to the QNM spectrum of Brans-Dicke black holes \cite{kwon86}, and hence the TeVeS QNMs also deviate by the same relation to the Brans-Dicke QNMs.  We have confirmed the result of equation (\ref{root2}) numerically in section \ref{Freq}.

Na\"ively, one may expect QNMs in TeVeS to tend exactly to the QNMs of GR in the limit as the theory's parameters tend to zero.  However, this expectation breaks down as the perturbation equation being considered herein does not exist in GR because of the lack of a vector field, and hence the derivation of the wave equation is not applicable.

\subsection{Wave equation in Isotropic Coordinates}\label{isopert}
Equation (\ref{timeevolSchw}) is not appropriate for studying details of QNMs due to the non-trivial coefficient preceding the time derivative.  Interestingly, performing a coordinate transformation from the Schwarzschild coordinates to the original isotropic coordinates defined by the transformation (\ref{SchwCoord}), together with the substitution
\begin{align}
    \Psi(t,r):={\rm e}^{\varphio}\left(1+\frac{M}{2r}\right)^{2}\Phi(t,r),
\end{align}
leads to the standard form of the wave equation
\begin{align}
    \left(\frac{\da^{2}}{\da t^{2}}-\frac{\da^{2}}{\da r_{\star}^{2}}+V\right)\Psi=0.\label{wave}
\end{align}
Here, we have introduced a new coordinate defined by
\begin{align}
    \frac{dr_{\star}}{dr}=\sqrt{2}{\rm e}^{2\varphi_{c}}\left(1+\frac{M}{2r}\right)^{3-2\delta_{\pm}}\left(1-\frac{M}{2r}\right)^{-1+2\delta_{\pm}},\label{coord}
\end{align}
which is a ``tortoise-like'' coordinate, in the sense that it maps $r\in\left(M/2,\infty\right)$ to $r_{\star}\in\left(-\infty, \infty\right)$, providing $-1+2\delta_{\pm}<0$.  Note that in isotropic coordinates the horizon occurs at $r=M/2$.

The potential, $V(r_{\star})$, is now given by the complicated expression
\begin{widetext}
\begin{align}
    V
    =&\frac{{\rm e}^{-4\varphi_{c}}}{2r^{2}}\left(1+\frac{M}{2r}\right)^{-2\left(3-2\delta_{\pm}\right)}\left(1-\frac{M}{2r}\right)^{2\left(1-2\delta_{\pm}\right)}\Bigg\{\ell\left(\ell+1\right)\notag\\
    &+\frac{M}{2r}\left[2-\left(1-\delta_{\pm}\right)\frac{M}{r}\right]\left[2\left(1-\delta_{\pm}\right)-\frac{M}{r}\right]\left(1+\frac{M}{2r}\right)^{-2}\left(1-\frac{M}{2r}\right)^{-2}\Bigg\}.\label{V}
\end{align}
\end{widetext}

Throughout section \ref{BHperts} we will look for QNMs of the above system of equations.  For these purposes it is convenient to assume a harmonic time dependence
\begin{align}
    \Psi(t,r)=\tilde{\Psi}(r){\rm e}^{i\omega t},
\end{align}
where $\om=\om_{{\rm Re}}+i\om_{{\rm Im}}$, which implies equation (\ref{wave}) becomes
\begin{align}
    \frac{d^{2}\tilde{\Psi}}{dr_{\star}^{2}}+\left(\om^{2}-V\right)\tilde{\Psi}=0. \label{SchrodEq}
\end{align}
In the above definition, $\om_{{\rm Re}}$ represents the frequency of the QNM while $\om_{{\rm Im}}$ is its damping time.  Any mode with $\om_{{\rm Im}}<0$ will grow exponentially in time, and hence represents an unstable mode.

\subsection{Behaviour of the potential}\label{potsec}
Determining stability and finding the QNM spectrum is now a one-dimensional scattering problem that is dependent solely on the form of the potential, $V$, defined in equation (\ref{V}).  It is therefore worth spending some time establishing various properties of this function.

We firstly note that the potential vanishes at spacelike infinity, independent of the value of $\varphi_{c}$.  The behaviour approaching the horizon, $r=M/2$, is more difficult as this is dependent on the value of $\delta_{\pm}$.  Bearing in mind that, for the black hole case, $\delta_{-}<0$ and $\delta_{+}>0$, we find

\begin{align}
    \lim_{r\rightarrow \frac{M}{2}^{+}}V(r)=\left\{\begin{array}{lcc}
        0 & {\rm for}  & \delta_{-}\\
        -\infty & {\rm for} & \delta_{+}
        \end{array}\right..
\end{align}

As discussed, the case of $\delta_{+}$ necessarily has regions of the spacetime that allows for superluminal propagation of scalar field perturbations.  While it has been heavily debated whether this is physically allowed to occur in nature (see for example \citet{ellis07}, \citet{bruneton07} and references therein), it is clear that this induces a pathology on the wave equation governing scalar perturbations in the form of a divergent potential function.  When we are discussing such superluminal perturbations we must be careful with our definition of ``horizon''.  Indeed in this case the surface at $r=M/2$ is a photon horizon associated with the null cones of the physical metric.  Superluminal propagation of scalar field perturbations implies that a perturbation at some $r<M/2$ could escape through the horizon at $r=M/2$ out to infinity.  We believe that this induces the pathological behaviour of the potential for the $\delta_{+}$ case, however we reserve further exploration of the effect of this to a future article.

\begin{figure}
    \begin{center}
    \includegraphics[width=0.47\textwidth]{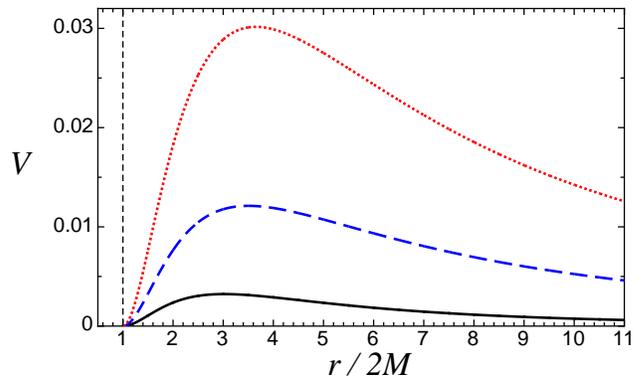}
    \caption{
    Potential, $V$, for the scalar perturbation wave equation for a black hole (i.e. $a=2$) with $\mathcal{K}=0.1$, $k=0.01$ and $\varphi_{c}=0.003$.  Note that $r$ is an isotropic radial coordinate implying the horizon of the spacetime is at $r=M/2$. This is the case of $\delta_{-}$, although at this range plots of $\delta_{+}$ look similar (the effect of the sign choice in $\delta_{\pm}$ becomes more relevant for higher $\ell$'s).  Here, the thick black, dashed blue and dotted red lines represent $\ell=0$, $1$ and $2$ respectively.
        }
    \label{fig:canonical}
\end{center}
\end{figure}

\begin{figure}
    \begin{center}
   \includegraphics[width=0.47\textwidth]{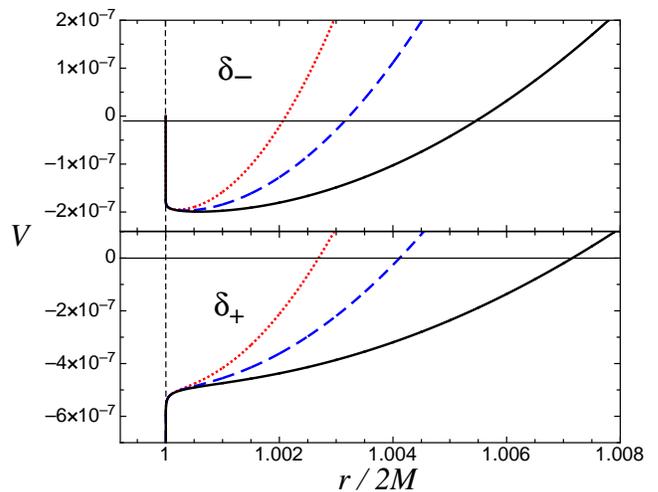}
    \caption{
    A zoomed in view of figure \ref{fig:canonical} for $\delta_{-}$ (top panel) and $\delta_{+}$ (bottom panel).  As stated in the text, as $r\rightarrow M/2$ the potential goes to zero for all $\ell$'s for the $\delta_{-}$ case, however diverges to negative infinity for the case of $\delta_{+}$.  This divergent behaviour is associated with the superluminal propagation of the scalar perturbations.
        }
    \label{zoom}
\end{center}
\end{figure}

It is also of interest to determine when the potential function is negative as such regions of a potential can represent bound states which imply growing modes.    Consider, for the moment, the $\ell=0$ case.  From equation (\ref{V}), one can show that for $\delta_{-}$ the potential is negative in the region
\begin{align}
    \frac{M}{2}\le r<\frac{M}{2\left(1-\delta_{-}\right)},
\end{align}
and for $\delta_{+}$ the potential is negative for
\begin{align}
    \frac{M}{2}<r<\frac{M\left(1-\delta_{+}\right)}{2}.
\end{align}
It is trivial to show that for all values of $\ell$ there exists a negative portion of the potential, however the largest negative region is realized for the $\ell=0$ case described above.  A generic plot of the potential is shown in figure \ref{fig:canonical}, for the $\delta_{-}$ case with $\ell=0$, $1$ and $2$.  We note that at this scale the difference in behaviour between the $\delta_{-}$ and $\delta_{+}$ cases can not be seen.  However, when one zooms into the region close to the horizon, figure \ref{zoom}, the difference in the respective functions is apparent.

\section{Results}\label{BHperts}
\subsection{Stability analysis
}\label{TimeEvol}
The presence of a single unstable mode can do severe damage to a theory as a viable alternative theory of
gravity.  The stability analysis of spherically symmetric black holes is generally rendered rather simple due to the property that any growing,
unstable mode, must necessarily be purely imaginary (\citet[for example see][]{konoplya10}).  That is, if an unstable mode exists
(i.e. such that $\om_{{\rm Im}}<0$), it must have zero real part, and therefore will not be oscillatory in nature.

\begin{figure}
    \begin{center}
    \includegraphics[width=\columnwidth]{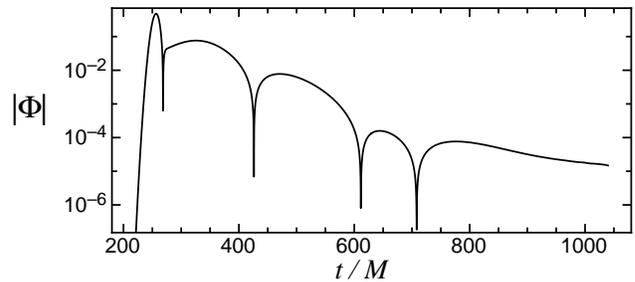}
    \caption{
      Typical temporal evolution of scalar field for $\ell=0$ with $\mathcal{K}=k=0.1$ and $\varphi_{c}=0.003$.  The presence of oscillations in the evolution implies non-zero $\om_{{\rm Re}}$ which, due to the spherically symmetric nature of the background spacetime, implies the mode is stable.  Due to the strong damping of the $\ell=0$ mode, only a few oscillations can be seen in the evolution.
    }
    \label{ell0}
\end{center}
\end{figure}

We have performed time evolutions of equation (\ref{timeevolSchw}) for various values of the spherical harmonic $\ell$.  The $\ell=0$ mode has the largest negative region in the potential, and hence is the most likely to be unstable.  In figure \ref{ell0} we plot the time evolution of a typical $\ell=0$ mode (the figure uses the example where $\mathcal{K}=k=0.1$ and $\varphi_{c}=0.003$).  Consistently with the case with scalar perturbations of the Schwarzschild black hole in GR, the $\ell=0$ mode has a large imaginary part (see section \ref{Freq}), implying the signal is significantly damped after only a few oscillations.  Whilst this makes it difficult to extract robust measurements of the frequencies and damping times for these modes, one can clearly see from figure \ref{ell0} that oscillations are present, implying these modes are stable.  As discussed above, larger values for the vector and scalar field coupling parameters act to increase the size of the negative region in the potential, which implies these cases are more likely to be unstable.  We have extended our numerical analysis for extreme values of the parameter space, including variations of $\varphi_{c}$, and find no evidence of any unstable modes.

\begin{figure}
    \begin{center}
    \includegraphics[width=\columnwidth]{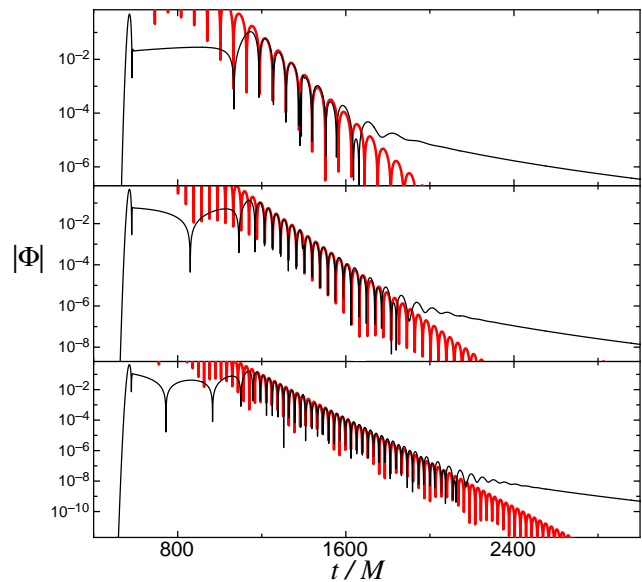}
    \caption{
     Temporal evolution of scalar field for $\ell=1$ (top panel), $\ell=2$ (middle) and $\ell=3$ (bottom) with $\mathcal{K}=k=0.1$ and $\varphi_{c}=0.003$.  The numerical time evolution is shown in black which includes the region of quasinormal ringing as well as the late-time tail.  Overlaid in red is the damped sinusoidal oscillations of the fundamental mode calculated using the WKB method (see section \ref{Freq}).
        }
    \label{manyell}
\end{center}
\end{figure}

In figure \ref{manyell} we plot the temporal evolution for the $\ell=1$, $2$ and $3$ modes for $\mathcal{K}=k=0.1$ and $\varphi_{c}=0.003$.  As expected there is no evidence of instabilities present for these, or any other values of the parameter space.  We have further calculated evolutions for higher $\ell$'s and again find no evidence of any instabilities for this type of perturbation.

Overlaid on figure \ref{manyell} are damped sinusoidal oscillations where the frequency and damping times are those found using the WKB method in the following section.  The WKB method is sensitive to the peak of the potential, whereas the time evolution accounts for the entire potential.  One can see from these figures that the WKB method gives extremely accurate results, implying the negative region of the potential contributes to the evolution of the perturbations on a level commensurate with the overall accuracy of the various schemes.

We have further verified the stability of the solutions using the time independent form of the wave equation (\ref{SchrodEq}). The method we used is
based on the fact that, if unstable modes exist ($\om_{{\rm Im}}<0$), the boundary conditions for the radial perturbation function $\tilde{\Psi}(r)$
become zero on both boundaries and the corresponding boundary value problem defined by equation (\ref{SchrodEq}) is
self-adjoint \cite{thorne78, doneva10b}.  This implies that, for the unstable modes, $\omega^{2}$ is real and negative and $\omega^{2}$ should be greater than the minimum of the potential, $V_{\rm min}$, for all of the bound states.  Thus, in order to find unstable modes, we integrate equation (\ref{SchrodEq}) with initial condition $\tilde{\Psi}|_{r_{\star}\rightarrow-\infty}=0$ for test values of $\omega^{2}$, starting from $\omega^{2}=V_{\rm min}$ to $\omega^{2}=0$.  An eigenfrequency is then found when the right boundary condition, $\tilde{\Psi}|_{r_* \rightarrow \infty}=0$, is fulfilled \cite{pryce93}.  When applying this method to our problem, it turns out that no unstable modes exist for all of the studied values of the parameters and the black holes are therefore stable against the considered perturbations.  


\subsection{Frequency domain calculations}\label{Freq}
Given that no unstable modes exist, the next step is to calculate the QNM modes governed by equation (\ref{SchrodEq}).  We have used two methods -- a direct integration (shooting) method and also the WKB method.  

The first method we used is the shooting method introduced by \citet{chandrasekhar75}
(see also \cite{kokkotas88} and \cite{doneva10b}). A strong aspect of this method is that it takes into account the presence of the negative minimum of the potential, however it is also prone to suffer from numerical instabilities due to this minimum and also the fact that the first derivative of the potential (\ref{V}) is divergent at the left boundary, $r=M/2$.
For this reason it was important to use an alternative method to confirm our results.
 As can be seen in figure \ref{fig:canonical} and the top panel of figure \ref{zoom}, the negative minimum of the potential
is extremely small compared to the positive maximum implying it is reasonable also to apply the WKB method \cite{schutz85,Iyer87a,kokkotas88,konoplya03}.

\begin{figure}
    \begin{center}
    \includegraphics[width=0.45\textwidth]{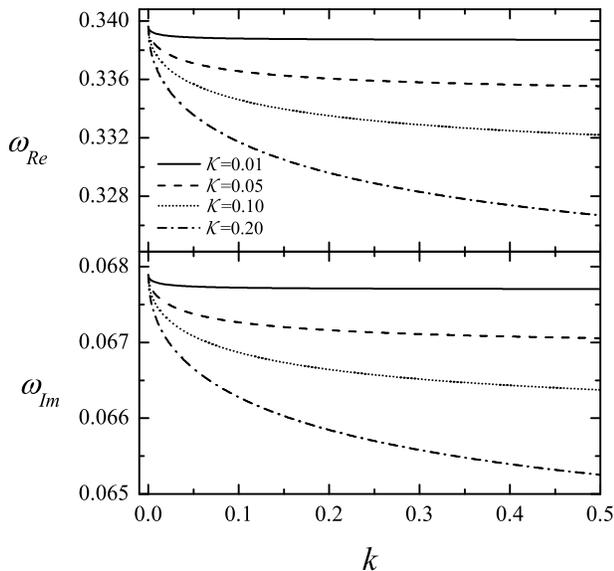}
    \caption{
    The real (top) and the imaginary (bottom) part of the fundamental $\ell=2$ QNM frequencies as a function of the scalar field coupling parameter $k$ for several values
    of the vector field coupling parameter $\mathcal{K}:=K+K_{+}-K_{4}$. The cosmological value of the scalar field is $\varphi_c=0.003$ and $\om_{{\rm Re}}$ and $\om_{{\rm Im}}$ are shown in units of $M$.
    }
    \label{fig:ReIm_ksmall}
\end{center}
\end{figure}

\begin{figure}
    \begin{center}
    \includegraphics[width=0.45\textwidth]{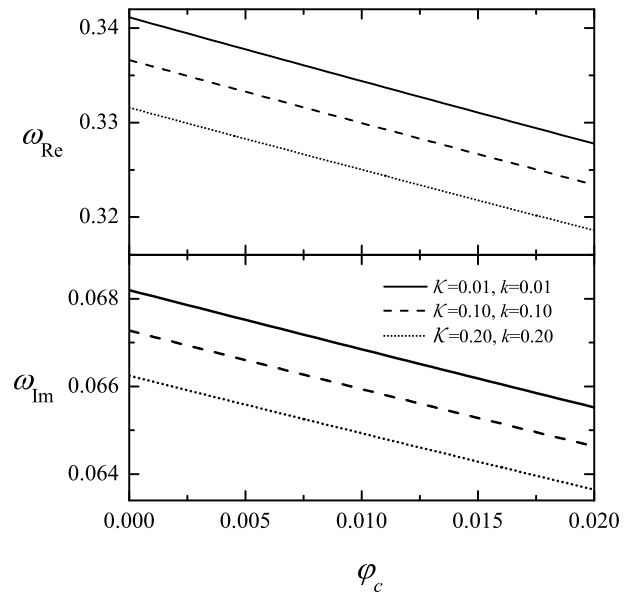}
    \caption{
    The real (top) and the imaginary (bottom) part of the fundamental $\ell=2$ QNM frequencies as a function of the cosmological value of the scalar field
    $\varphi_c$ for several values of the parameters $\mathcal{K}$ and $k$.  Here, $\om_{{\rm Re}}$ and $\om_{{\rm Im}}$ are shown in units of $M$
    }
    \label{fig:ReIm_kphic}
\end{center}
\end{figure}

In figures \ref{fig:ReIm_ksmall} and \ref{fig:ReIm_kphic} we show the fundamental ($n=0$) $\ell=2$ QNM frequencies and damping times obtained from the WKB method for various values of the theory's coupling parameters\footnote{For physically relevant values of the theory's parameters, the results obtained from the WKB and the shooting method
differ by up to 1\% and the difference becomes smaller for smaller values of the coupling parameters.}. As can be seen for relevant values of the parameters the frequencies only vary by a few percent throughout the range of the physically relevant parameter space.  We have displayed this in figures \ref{fig:ReIm_ksmall} and \ref{fig:ReIm_kphic} only for the $\ell=2$ mode however, as the associated errors for the $\ell=0$ and $\ell=1$ modes are comparable to the change in frequencies associated with the variation of the parameter space.  Nevertheless, our calculations have shown that the qualitative behaviour of the $\ell=0$ and $\ell=1$ modes are similar to the shown $\ell=2$ case.

Finally, from our calculations we have identified the limiting behaviour as the theory's various parameters go to zero which is shown in table \ref{table}.  The final column of table \ref{table} shows the corresponding QNMs for a Schwarzschild black hole in GR.  One can see that the errors produced are extremely small, even for the difficult to calculate $\ell=0$ case.  These numerical results confirm the discussion of section \ref{wavesec} and particularly the relation between GR and the limiting behaviour of TeVeS elucidated in equation (\ref{root2}).

\begin{table}
	\begin{tabular}{|c|c|c|}
	\hline
		$\ell$ &$ \om$ & $\om_{s}/\sqrt{2}$\\
		\hline
		$0$ & $0.0797+i0.0749$ & $0.0781+i0.0742$\\
		$1$ & $0.2071+i0.0691$ & $0.2071+i0.0691$\\
		$2$ & $0.3422+i0.0684$ & $0.3420+i0.0684$\\
		$3$ & $0.4779+i0.0682$ & $0.4776+i0.0682$\\
		\hline
	\end{tabular}
	\caption{\label{table}Fundamental quasinormal modes for the first three spherical harmonics in the limit as the theory's various coupling parameters tend to zero, $\om$ and for a Schwarzschild black hole in GR, $\om_{s}$.  The Schwarzschild QNMs in GR are calculated using the continued-fraction method, while $\om$ is calculated using the shooting method which has an error in this limit of a few percent for the $\ell=0$ mode and significantly smaller for the higher modes.  These numerical results confirm the relation in equation (\ref{root2}).}
\end{table}

\section{Conclusions}
We have calculated here the stability and QNM spectrum for spherically symmetric black holes in TeVeS and its generalization under perturbations of the scalar field intrinsic to the theory.  Despite the existence of a small negative region in the potential of the wave equation, we have shown that these black holes are generically stable to such perturbations.  In the limit as the various coupling parameters of the theory tend to zero, when one na\"ively expects to recover GR, we have shown that the QNM spectrum tends to a value different to the scalar perturbations of Schwarzschild black holes in GR and in Brans-Dicke theory by a factor of $\sqrt{2}$.  The reason for this is clear: as one takes the limit of small coupling parameters the background vector field does not vanish.  It is not until the parameters are set to zero that the background field vanishes, at which point the perturbation equation that we have explored (\ref{scalarwave}) is no longer a valid equation. In other words, this type of specific perturbation does not exist in General Relativity.

We have explored herein only a small branch of existing solutions in the strong-field regime in TeVeS.  Indeed, even within the subset of spherically symmetric solutions with temporal pointing vector fields we have all but ignored a broad range of exotic solutions that involve either superluminal propagation of scalar perturbations, naked singularities or both.  For these cases we have shown that the potential in the wave equation governing the perturbations generally (although there are some exceptions -- see the Appendix) diverges at the critical horizon representing either the naked singularity or the photon event horizon.  The case that allows for superluminal perturbations is severely complicated by the fact that perturbations appearing within the photon horizon can escape to future null infinity.  In fact, in these cases it is still not clear whether the central essential singularity (i.e. at $r=0$) is naked to scalar perturbations in the sense that these perturbations at $r=0$ can evolve to future null infinity.  We plan to explore the stability of these objects as well as those with naked singularities at the critical radius (i.e. those spacetimes with $a\neq2$) in future work.

\appendix*
\section{The wave equation for naked singularities}
In this appendix we show the more general wave equation for arbitrary values of the metric parameter $a$, which is therefore relevant for perturbations of naked singularities amongst other things.  Following the procedure outlined in section \ref{wavesec}, with an arbitrary $a$, leads to the wave equation (\ref{wave}), where the potential is now generalized to
\begin{align}
    V
    &=\frac{{\rm e}^{-4\varphi_{c}}}{2r^{2}}\left(1+\frac{M}{2r}\right)^{-2\left(1+\chi\right)}\left(1-\frac{M}{2r}\right)^{-2\left(1-\chi\right)}\Bigg[\ell\left(\ell+1\right)\notag\\
    &+\frac{M}{2r}\left(2-\frac{\chi M}{2r}\right)\left(\chi-\frac{M}{r}\right)\left(1+\frac{M}{2r}\right)^{-2}\left(1-\frac{M}{2r}\right)^{-2}\Bigg],\label{VApp}
\end{align}
where $\chi:=a-2\de_{\pm}$.  Moreover, the coordinate, $r_{\star}$ is now given by
\begin{align}
    \frac{dr_{\star}}{dr}=\sqrt{2}{\rm e}^{2\varphi_{c}}\left(1+\frac{M}{2r}\right)^{1+\chi}\left(1-\frac{M}{2r}\right)^{1-\chi}.\label{coordApp}
\end{align}

As with the more specific black hole case discussed in section \ref{potsec} the potential vanishes at spacelike infinity, however has the following limiting behaviour as one approaches the horizon
\begin{align}
    \lim_{r\rightarrow r_{c}^{+}}V(r)=\left\{\begin{array}{lcc}
        0 & {\rm for}  & \chi\ge 2\\
        -\infty & {\rm for} & \chi< 2
        \end{array}\right..
\end{align}
It is known from section \ref{sssvac} that $\delta_{+}>0$ for all values of $a$, $\cal{K}$ and $k$, while $\delta_{-}<0$ for certain values of $a$ and $\cal{K}$ depending on the relation given in (\ref{ineqaK}).

Again, independent of the behaviour at $r=M/2$, the potential necessarily goes negative for some region of the spacetime, apart from the exceptional case of $\chi=2$.  For the $\ell=0$ case, the potential has the following behaviour dependent on $\chi$:
\begin{align}
    V(r;\,\chi<2)<0\qquad&\forall\qquad \frac{M}{2}\le r<\frac{M}{\chi},\\
    V(r;\,\chi=2)>0\qquad&\forall\qquad r,\\
    V(r;\,\chi>2)<0\qquad&\forall\qquad \frac{M}{2}<r<\frac{\chi M}{4}.
\end{align}
That is, for all cases excluding that where $\chi=2$, the potential is negative for the immediate region of the spacetime beyond $r=M/2$.  At some finite value (either $M/\chi$ for $\chi<2$ or $\chi M/4$ for $\chi>2$) the potential is exactly zero, and is positive for all larger values of the radius.  It is trivial to show that for all values of $\ell$ there exists a negative portion of the potential, however the mode that is most likely unstable is the fundamental $\ell=0$ mode.  We leave the stability study of this broader class of spacetimes to future work.

\acknowledgments{We are indebted to Dimitrios Giannios, Kostas Glampedakis, Kostas Kokkotas, Roman Konoplya, and Stoytcho Yazadjiev for valuable
discussions. P.L. is supported by the Alexander von Humboldt Foundation and the Transregio 7 ``Gravitational Wave Astronomy,'' financed by the
Deutsche Forschungsgemeinschaft DFG (German Research Foundation). D.D. would like to thank the DAAD for support and the Institute
for Astronomy and Astrophysics T\"{u}bingen for its kind hospitality. D.D. was partially supported by the Bulgarian National Science Fund under Grants DO 02-257, VUF-201/06 and by Sofia
University Research Fund under Grant No 101/2010.  This work was also supported by an IKY-DAAD grant and we appreciate useful interactions with Th. Apostolatos.  We thank the anonymous referee for valuable comments regarding the manuscript.}

\bibliography{Bib}
\end{document}